\documentclass[aip, apl, reprint]{revtex4-1}

\usepackage[utf8x]{inputenc}

\usepackage{amsmath}
\usepackage{graphicx}
\usepackage{bm} %bold math
\usepackage{dcolumn} % Align table columns on decimal point
\usepackage[colorlinks=true, citecolor=blue, linkcolor=blue]{hyperref}%
\newcommand{\cor}[1]{{\color{black} #1}}

\draft % marks overfull lines with a black rule on the right

\begin{document}

%\title{Impurity-activated photoinduced modulation of terahertz waves in GaAs crystal} %Title of paper
\title{Impurity-induced modulation of terahertz waves in optically excited GaAs} %Title of paper

\author{A. S. Kurdyubov}

\author{A. V. Trifonov}

\author{I. Ya. Gerlovin}

\author{I. V. Ignatiev}
\affiliation{Spin Optics Laboratory, St. Petersburg State University, 1 Ul’anovskaya, Peterhof, St. Petersburg 198504, Russia}

\author{A. V. Kavokin}
\affiliation{Spin Optics Laboratory, St. Petersburg State University, 1 Ul’anovskaya, Peterhof, St. Petersburg 198504,
Russia}
\affiliation{CNR-SPIN, Viale del Politecnico 1, I-00133, Rome, Italy}
\affiliation{School of Physics and Astronomy, University of Southampton, SO17 1NJ Southampton, United Kingdom}

\date{\today}

\begin{abstract}
The effect of the photoinduced absorption of terahertz (THz) radiation in a semi-insulating GaAs crystal is studied by the pulsed THz transmission spectroscopy. We found that a broad-band modulation of THz radiation may be induced by a low-power optical excitation in the spectral range of the impurity absorption band in GaAs. The measured modulation \cor{factor} achieves 80\%. The amplitude and frequency characteristics of the resulting THz modulator are critically dependent on the carrier density and relaxation dynamics in the conduction band of GaAs. In semi-insulating GaAs crystals, the carrier density created by the impurity excitation is controlled by the rate of their relaxation to the impurity centers. The relaxation rate and, consequently, the frequency characteristics of the modulator can be optimized by an appropriate choice of the impurities and their concentrations. The modulation parameters can be also controlled by the crystal temperature and by the power and photon energy of the optical excitation. These experiments pave the way to the low-power fast optically-controlled THz modulation, imaging, and beam steering.
\end{abstract}


\maketitle %\maketitle must follow title, authors, abstract and \pacs

The electromagnetic radiation belonging to the terahertz spectral range is important field of active research promising for a wide range of applications~\cite{Tonouchi-NPot2007, Zhang-book2010, Chattopadhyay-IEEE2011, Belkin-PhysScr2015, Nagatsuma-NPhot2016}. Ability of THz radiation to penetrate most part of non-conductive materials makes it a promising tool for a non-destructive control of organic materials~\cite{Rutz-Waves2006}, real-time THz imaging~\cite{Busch-OptLett2012, Shams-IEEE2017}, and security applications~\cite{Jepsen-Laser2011, Cooper-IEEE2014}. Very broad band of THz radiation can be used for the ultrafast THz communications and data processing~\cite{Nagatsuma-NPhot2016, Jastrow-ElLett2008, Song-IEEE2011}. Realization of these applications requires, in particular, compact and reliable THz modulators. 

Modulators of electromagnetic waves in the optical spectral range are well developed by now. These devices are able to control the amplitude, phase, polarization, and other characteristics of optical waves. Realization of similar modulators for THz radiation is a challenging problem because of the much longer THz wavelength, which would require either the use of bulky macroscopic modulators or the use of new physical principles of modulation in compact solid state devices. One of the promising ways to realize a THz modulator is considered to be a photo-induced modification of the THz characteristics of semiconductor media by an optical excitation. 

Several modulation methods based on the optically-induced physical processes in semiconductors are proposed so far~\cite{Libon-APL2000, Okada-SciRep2011, Busch-OptLett2012, Cheng-OptExpr2013, Rahm-Waves2013, Steinbusch-OptExpr2014, Shams-ElectrLett2014}. In particular, pure Si and GaAs crystals have been used for this purpose~\cite{Cheng-OptExpr2013}. The interband optical excitation of the crystals creates free electrons and holes, which efficiently absorb the THz radiation. Due to the fast electron-hole recombination in the direct-gap semiconductors, e.g., GaAs, such modulators are characterized by the high operation rate (the switching time is of the order of units of ns)~\cite{Rahm-Waves2013}. At the same time, the fast recombination hampers the accumulation of the carriers and, consequently, the deep modulation of THz absorption. In the indirect-gap semiconductors, e.g., silicon, or in the type II quantum wells~\cite{Libon-APL2000}, the carrier lifetime is much larger and the required carrier density is easier to accumulate compared to the direct band gap systems~\cite{Cheng-OptExpr2013, Nozokido-RIKEN1995, Shams-IEEE2017}. However, the indirect-gap nature of this material requires relatively strong excitation power (Watts per cm$^2$) to achieve a noticeable THz absorption. Besides, the operation rate is of the order of tens of kHz limited by the carrier lifetime. 

In Ref.~\cite{Kurdyubov-FTT2017} it was found that, in a semi-insulating (SI) GaAs crystal, a significant concentration of the free carriers in the conduction band can be created by an optical excitation of the electrons bound to acceptor centers. The relaxation rate of the photo-excited electrons strongly depends on the binding energy of electrons at these centers, $E_a$. At \cor{low} temperatures, the corresponding relaxation time varies from fractions of microsecond for shallow centers with $E_a$ of order of several tens of meV to fractions of millisecond for much deeper centers~\cite{Dumke-PR1963}. The recombination is accelerated at the temperature rise. So, a variation of the type of centers to be optically excited and the crystal temperature allows one, in principle, to strongly vary the operation rate of a THz modulator. 

In the present work we study the spectral, amplitude, and frequency responses of a THz modulator based on a SI GaAs crystal at different temperatures. We demonstrate the evidence of the giant photomodulation effect in the spectral range corresponding to the impurity bands in SI GaAs crystals.
Samples under study are the thin plates of SI GaAs produced as substrates for epitaxial technologies. The THz experiments were performed by the use of a standard THz time-domain spectroscopy technique~\cite{Zhang-book2010}. The THz pulses were focused on the end face of the GaAs plate, which length along the THz propagation is of about few mm. This geometry of experiments allowed us to optimize the length of THz absorbing material. The transmitted THz radiation was detected by a system measuring the temporal profile of electromagnetic field of the THz pulses. A Fourier transformation of the profile gives the spectrum of the transmitted pulses. The THz part of the setup, from the THz source up to the THz detector, was placed into a dry box to reduce the THz absorption by the water vapor.

To induce the THz absorption, the front surface of the plate was illuminated by radiation of a continuous-wave (CW) tunable laser. Variation of the laser spot size on the sample allowed us to vary the length of the THz absorbing volume. We studied the photo-induced THz absorption as a function of the photon energy and power of the CW excitation and of the sample temperature. To study the frequency dependence of the absorption, we modulated the laser radiation by an acousto-optical modulator in the frequency range from 5~kHz to 1 MHz.   

\begin{figure}
\includegraphics[width=0.99\columnwidth]{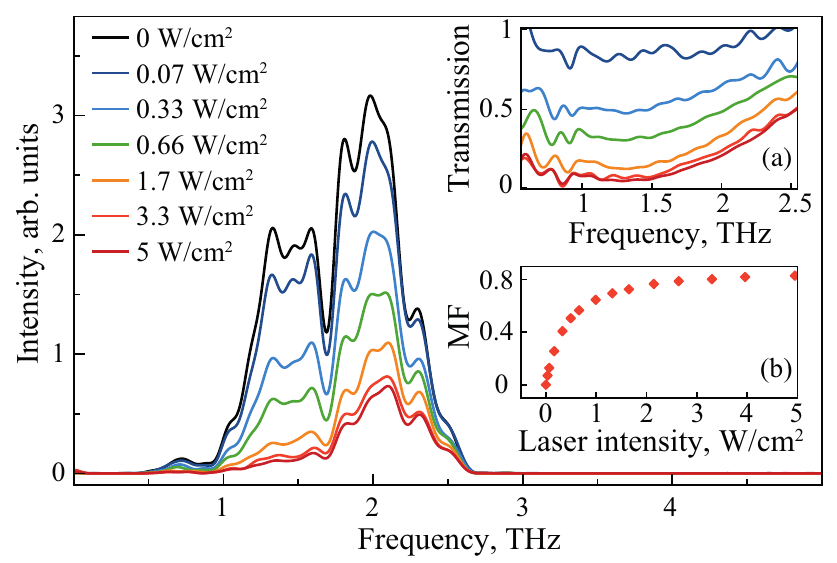}%
\caption{Fourier spectra of the THz radiation transmitted through the SI GaAs at different excitation densities with photon energy, $E_{\text{CW}}=1498$~meV. The laser spot diameter along the THz passway, $D = 2$~mm.   The upper inset shows the THz transmission spectra normalized to the reference spectrum measured with no optical excitation. The low inset shows the power dependence of the modulation factor calculated using Eq.~(\ref{eq1}). The sample temperature $T = 4$~K.}
\label{Fig1}%
\end{figure}

Fig.~\ref{Fig1} shows the spectra of THz pulses transmitted through the sample illuminated by the CW radiation with different excitation densities. Each spectrum is a broad band of about 1.5~THz width centered at 1.7~THz. The spectral dip observed at the central frequency is caused by absorption of THz radiation in the box by the residual water vapor. The CW excitation is found to result in a decrease of the transmission of the THz radiation, which is uniform over the whole THz spectrum. The transmission gradually decreases with the increase of the excitation power as it is shown in the insets of Fig.~\ref{Fig1}. However, at some excitation density, the photo-induced THz absorption is saturated, see low inset in Fig.~\ref{Fig1}.

 The effect of the photo-induced THz absorption can be characterized by a modulation factor (MF) defined as\cite{Wang-APL2017,Matsui-OptLett2013}
\begin{equation}
\label{eq1}
MF=1-\frac{\int E^2_{\text{CW-on}}d\omega}{\int E^2_{\text{CW-off}}d\omega},
\end{equation}
where $E$ is the electric field of the THz wave measured in the presence ($E_{\text{CW-on}}$) and absence ($E_{\text{CW-off}}$) of the CW excitation. As seen in the low inset of Fig.~\ref{Fig1}, the \cor{MF} can reaches 80\% at \cor{relatively} low excitation density of about 1~W/cm$^2$.

\begin{figure}
 \includegraphics[width=0.99\columnwidth]{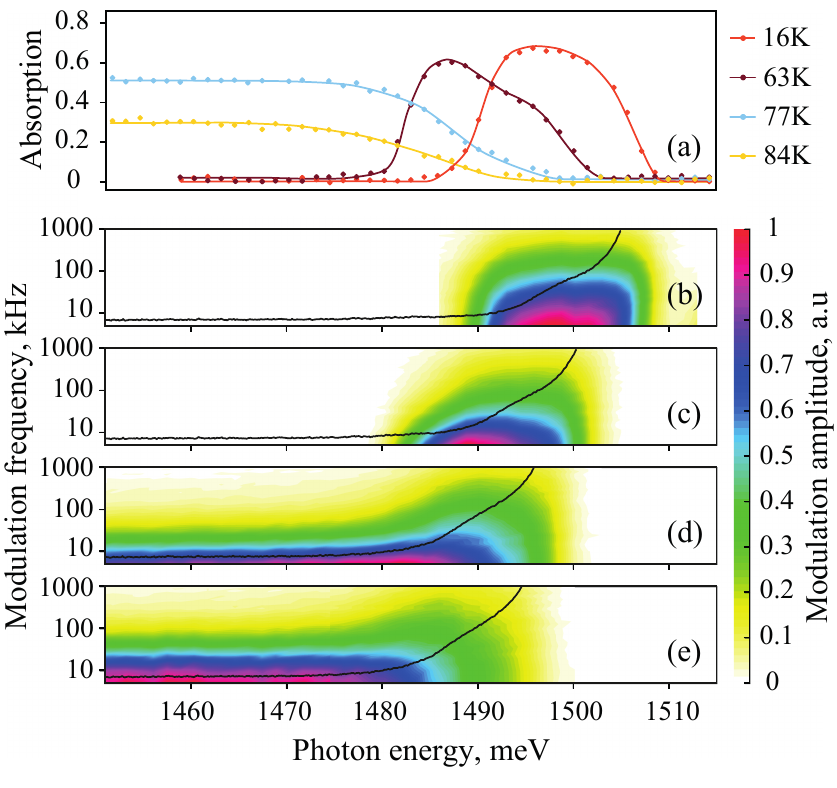}%
 \caption{(a) The photoexcitation spectra of THz absorption (action spectra) in SI GaAs at different sample temperatures. The excitation density is $\approx 2$~W/cm$^2$. (b-e) Relative modulation amplitude of the THz radiation as a function of the photon energy and modulation frequency of optical excitation at temperatures $T = 16$~K (b), $T = 63$~K (c), $T = 77$~K (d), $T = 84$~K (e). The color bar shows the scale of the THz modulation amplitude normalize\cor{d} at its maximal value for each temperature. The optical excitation density $\approx 3$~W/cm$^2$. The black curves show the spectral dependences of optical density of the sample under study at respective temperatures.} 
 \label{Fig2}
 \end{figure}

At the low sample temperature, the optical excitation spectrum of the THz absorption, that is the dependence of the THz absorption coefficient on the photon energy of the excitation, is characterized by a relatively narrow spectral band centered at 20~meV below the band gap of GaAs [see Fig.~\ref{Fig2}(a)]. The excitation above this band, in particular into the spectral range of the fundamental absorption of GaAs, does not affect the transmitted THz radiation. This unusual, at the first glance, result indicates that such optical excitation does not create any noticeable carrier density in spite of the large absorption of the optical radiation. 

An increase of the sample temperature is followed by a shift of the excitation spectrum of the THz absorption (referred hereafter to as the ``action band I'') to the lower photon energies. This shift is similar to that of the GaAs band gap. The magnitude of the photo-induced THz absorption in the maximum of the excitation band decreases with the temperature rise. At $T > 80$~K, this band virtually disappears. 
The energy position of the action band I and its temperature variation points out that it is related to the photo-ionization of the negatively charged acceptor centers~\cite{Sze-1968, Schultz-model2009}. The 20-meV energy shift of this band relative to the fundamental absorption edge corresponds to the acceptors created by the carbon impurities~\cite{Heilman-SST1990}. These are the most common impurities related to peculiarities of the GaAs growth technology.

At the larger sample temperatures, $T > 77$~K, another mechanism of the photo-induced absorption appears. Its efficiency is virtually constant in the spectral range of 1450 -- 1480~meV. The further rise of the sample temperature is followed by a gradual decrease of the THz absorption. We suggest that the action band II observed at these temperatures is related to the deeper acceptors, whose energy is spread out over a wide range~\cite{Sze-1968}. For some samples of GaAs studied by us, the action band II of the photo-induced THz absorption has been observed even at the low sample temperatures of about 4~K. We assume that this effect is caused by the much larger density of the deep acceptors in these samples. A discussion of these differences in the spectral behavior of different samples of GaAs grown by different technologies is out of scope of the present paper.

For the study of temporal characteristics of the photo-induced THz absorption, we have measured the dependence of the \cor{modulation amplitude of the THz radiation} on the amplitude modulation frequency of the CW excitation. \cor{The modulation amplitude is the difference between the THz signals when the CW excitation is switched on and off normalized to the same difference measured at low modulation frequency.} A general behavior of the modulation effect is illustrated in Fig.~\ref{Fig2}(b-e). One can see that the increase of the modulation frequency gives rise to a decrease of the THz \cor{modulation amplitude}. At the relatively low temperatures, $T < 20$~K, when only the action band I is observed, the THz \cor{modulation amplitude} becomes twice lower at the frequency of about 100~kHz. At $T \ge 77$~K where the action band II is observed, similar suppression of the THz modulation occurs at the considerably smaller frequencies [see Fig.~\ref{Fig2}(d,e)].  

To quantitatively describe the frequency dependences of the THz modulation, we have simulated the processes of the excitation and relaxation of the free carriers induced by the amplitude modulation of the CW laser beam forming a meander with period $T$. It is assumed that the relaxation of free carriers responsible for the THz absorption is a single-exponential decay with characteristic time $\tau$. The temporal dependence of the carrier density and, consequently, of the THz modulation amplitude is described by:  
\begin{equation}
\label{eq.FR}
F(t)=F_0\times
\left\{\begin{array}{rl}
1-\frac{2e^{-t/\tau}}{1+e^{-T/2\tau}},&~\text{for } 0<t<T/2,\\
\frac{2e^{-(t-T/2)/\tau}}{1+e^{-T/2\tau}}-1,&~\text{for } T/2<t<T.
\end{array}\right.
\end{equation}
Here $F_0$ is the modulation amplitude at the very slow modulation, $T \gg \tau$. Deriving Eqs.~(\ref{eq.FR}) we have assumed that the THz absorption amplitude is proportional to the carrier density in the conduction band. 

The THz signal is measured by a system including a lock-in amplifier so that the detected signal is:
\begin{eqnarray}
\label{eq.FRx}
&X(\nu)=\frac{1}{T}\int F(t)\cos(2\pi\nu t)dt, &\\ \nonumber
&Y(\nu)=\frac{1}{T}\int F(t)\sin(2\pi\nu t)dt, &\\ \nonumber
&R(\nu)=\sqrt{X(\nu)^2+Y(\nu)^2},&
\end{eqnarray}
where $\nu = 1/T$ is the modulation frequency, $X(\nu)$ and $Y(\nu)$ are the signals detected in the two channels with the $\pi/2$ phase shift; $R(\nu)$ is the amplitude of the signal. The integration is fulfilled over the modulation period $T$.

\begin{figure}
\includegraphics[width=0.99\columnwidth]{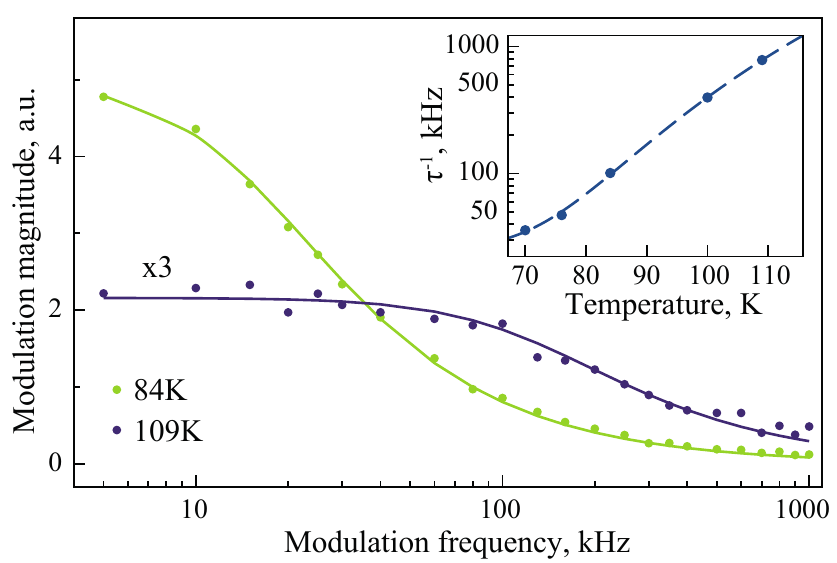}%
\caption{Amplitude-frequency characteristics, $R(\nu)$, of the THz modulator measured at two temperatures under the optical excitation into the action band II ($E_{\text{exc}} = 1450$~meV). Points are the experimental data, solid lines are the results of modelling by Eqs.~(\ref{eq.FR}, \ref{eq.FRx}). Inset shows the temperature dependence of the relaxation rate (points) obtained from the fits of the amplitude-frequency characteristics. Dashed line is the fit by Eq.~(\ref{eq.bose}). Note the logarithmic scale for $\tau^{-1}$.}
\label{Fig3}
\end{figure}

The frequency dependencies of the THz modulation obtained at temperatures $T > 80$~K for the action band II are well reproduced by Eqs.~(\ref{eq.FR},\ref{eq.FRx}) as it is shown in Fig.~\ref{Fig3}. The obtained agreement with the experiment allows us to estimate the corresponding relaxation rate, $1/\tau$. We found that this rate rapidly rises with the sample temperature, see inset in Fig.~\ref{Fig3}.
The physical origin of this strong acceleration of the relaxation can be in the thermally-activated population of the deep centers with electrons coming from the valence band. As a result, the electrons optically excited to the conduction band may recombine with the holes appearing in the valence band. This process is much faster than the electron recombination with the holes at the deep centers. The thermally-induced increase of the relaxation rate is well described by an activation function: 
\begin{equation}
\label{eq.bose}
\tau^{-1}=\tau_0^{-1}+\frac{\tau_1^{-1}}{\exp(\Delta E/kT)-1},
\end{equation}
where $\tau_0^{-1}$, $\tau_1^{-1}$, and $\Delta E$ are the fitting parameters. As seen in the inset of Fig.~\ref{Fig3}, this function describes well the experimental data with activation energy $\Delta E = 74$~meV corresponding to the energy of two LO phonons in GaAs~\cite{Strauch-JPhys1990}.
 
The similar study of the action band I observed at the lower temperatures ($T < 80$~K) shows that its amplitude-frequency characteristics cannot be simulated by Eqs.~(\ref{eq.FR}, \ref{eq.FRx}). This means that the electron relaxation to the shallow centers cannot be described by a single exponential law. A cutoff frequency corresponding to the twofold decrease of the THz absorption is of about 100~kHz and weakly depends on the sample temperature.

The obtained results demonstrate that the THz absorption in the SI GaAs crystals can be initiated by a relatively weak optical excitation in the spectral range of the impurity absorption bands. This excitation is more efficient from point of view of the THz absorption than the excitation within the spectral range of the fundamental GaAs absorption in spite of the much lower optical density of the impurity bands. This unusual, at the first glance, effect is \cor{due to large} lifetime of the electrons excited from the impurity centers that is beneficial for electron accumulation in the conduction band. In the case of the interband excitation, fast electron-hole recombination (of the order of ns) prevents the electron accumulation and, consequently, a noticeable THz absorption.

The weak THz absorption in the case of the interband excitation would be, in principle, related to the very thin near-surface layer of the order of 1~$\mu$m where electrons in the conduction band are created. The THz radiation propagating along the crystal plate may not ``feel'' this layer. To verify the origin of the weak THz absorption, we have changed the geometry of the experiment and directed the THz radiation along the optical excitation beam that is perpendicular to the GaAs plate. In this case, the THz waves pass through the optically excited layer and the photocreated electrons may absorb the THz radiation. This experiment has shown that the THz absorption is weak, as before, in the case of the interband excitation and it is much stronger in the case of the optical excitation into the impurity band. So, this experiment confirms that the main reason for the low THz absorption in the case of the interband optical excitation is the fast electron-hole recombination.

The lifetime of the electrons created by the photo-ionization of impurity centers is determined by the reverse relaxation of electrons back to the impurities. The rate of this relaxation depends on the density of the impurity centers and on their binding energy. The relatively slow relaxation of electrons to the impurity centers at low temperatures limits the THz modulation rate to about 100~kHz in the samples studied. 

At the elevated crystal temperatures, the deep impurity centers can be populated with electrons due to the thermally-activated ionization of the shallow centers. This process explains the appearance of the action band II and the disappearance of the action band I. In turn, the shallow centers can be repopulated with electrons that leads to the generation of holes in the valence band. The fast electron-hole recombination strongly accelerates the electron relaxation that allows one to control the cutoff frequency of the THz modulation in a wide range, as it is shown in the inset of Fig.~\ref{Fig3}. In the same time, the fast recombination reduces the electron density and, consequently, the THz absorption and the fixed optical excitation power. 

The experimental study of the main characteristics of the impurity-related photoinduced THz absorption in SI GaAs shows that there are real prospects for creating the broad-band THz modulators on the basis of this effect. The cutoff modulation frequency achievable for such modulators is of the order of several tens of kHz at the liquid nitrogen temperature and it rapidly increases up to 100 kHz and more with the crystal temperature rise. The required optical excitation power is a fraction of Watt and the wavelength (of about 850~nm) falls within the operating range of cheap, commercially available, semiconductor lasers. It is essential that the excitation in the spectral range of the impurity band allows one to reach nearly the 100\% \cor{MF} of the THz radiation by the use of GaAs crystal plates with thickness of several mm. The growth of the GaAs crystals required for the modulators is the well developed area of industry because SI GaAs crystal plates are widely used as substrates for the epitaxial growth of heterostructures. 

The effect of the impurity-induced THz absorption at the optical excitation allows one to vary the modulation parameters in the wide range. In particular, the frequency characteristics of the modulators, which is limited by the relaxation rate of electrons to the impurity centers, can be optimized by the appropriate choice of the type and density of the impurities. The reduction of the relaxation time broadens the modulation frequency band but it reduces, other parameters being equal, the density of photocreated carriers controlling the THz \cor{MF}. To improve the \cor{MF} without degrading the frequency response, one can use the stronger optical excitation and/or the larger crystal thickness along the THz propagation path. Additional methods of control of the modulation parameters are the variation of the crystal temperature and of the spectrum of the excitation light.

An important feature of such modulators is the opportunity of employing them as the dynamic two-dimensional amplitude filters, which are required for the parallel data transmission systems operating in the THz spectral range~\cite{Shams-ElectrLett2014} as well as for the THz beam steering~\cite{Shams-IEEE2017} by means of the dynamic holograms, Fresnel optics etc.

The authors thank the Resource Center ``Nanophotonics'' St. Petersburg State University for the technical support of the experiments. This work was supported by the Russian Foundation for Basic Research, project No. 15-59-30406 RT.


\begin{thebibliography}{50}

\bibitem{Tonouchi-NPot2007} M. Tonouchi, Nature Photon. {\bf 1}, 97 (2007).
\bibitem{Zhang-book2010} X.-C. Zhang and J. Xu, {\em Introduction to THz Wave Photonics} (Springer Science and Business Media, LLC 2010).
\bibitem{Chattopadhyay-IEEE2011} G. Chattopadhyay, IEEE Trans. THz Sci. Techn. {\bf 1}, 33 (2011). 
\bibitem{Belkin-PhysScr2015} M. A Belkin and F. Capasso, Phys. Scr. {\bf 90}, 118002 (2015).
\bibitem{Nagatsuma-NPhot2016} T. Nagatsuma, G. Ducournau, and C. C. Renaud, Nat. Phot. {\bf 10}, 371 (2016).
\bibitem{Rutz-Waves2006} F. Rutz, M. Koch, S. Khare, M. Moneke, H. Richter, and U. Ewert, Int. Journal Infrared and Millimeter Waves {\bf 27}, 547 (2006).
\bibitem{Busch-OptLett2012} S. Busch, B. Scherger, M. Scheller, and M. Koch, Opt. Lett. {\bf 37}, 1391 (2012).
\bibitem{Shams-IEEE2017} M. I. B. Shams, Z. Jiang, S. M. Rahman, L.-J. Cheng, J. L. Hesler, P. Fay, and L. Liu, IEEE Transact. THz Sci. Techn. {\bf 7}, 310 (2017).
\bibitem{Jepsen-Laser2011} P. U. Jepsen, D. G. Cooke, and M. Koch,  Laser Photonics Rev. {\bf 5}, 124 (2011).
\bibitem{Cooper-IEEE2014} K. B. Cooper and G. Chattopadhyay, IEEE Microwave Mag. {\bf 15}, 51 (2014).
\bibitem{Jastrow-ElLett2008} C. Jastrow, K. Munter, R. Piesiewicz, T. Kurner, M. Koch, and T. Kleine-Ostmann, Electron. Lett. {\bf 44}, 213 (2008)
\bibitem{Song-IEEE2011} H.-J. Song and T. Nagatsuma, IEEE Trans. THz Sci. Techn. {\bf 1}, 256 (2011).
\bibitem{Libon-APL2000} I. H. Libon, S. Baumgärtner, M. Hempel, N. E. Hecker, J. Feldmann, M. Koch, P. Dawson, Appl. Phys. Lett. {\bf 76}, 2821 (2000).
\bibitem{Okada-SciRep2011} T. Okada and K. Tanaka, Scientific Reports {\bf 1}, 121 (2011).
\bibitem{Cheng-OptExpr2013} L.-J. Cheng and L. Liu, Optics Express {\bf 21}, 028657 (2013).
\bibitem{Rahm-Waves2013} M. Rahm, J.-S. Li, W. J. Padilla, J. Infrared Milli Terahertz Waves {\bf 34}, 1 (2013).
\bibitem{Steinbusch-OptExpr2014} T. P. Steinbusch, H. K. Tyagi, M. C. Schaafsma, G. Georgiou, and J. Gomez Rivas, Opt. Expr. {\bf 22}, 26559 (2014).
\bibitem{Shams-ElectrLett2014} M. I. B. Shams et al., Electron. Lett. {\bf 50}, 801 (2014).
\bibitem{Nozokido-RIKEN1995} T. Nozokido, H. Minamide, and K. Mizuno, RIKEN Review {\bf11}, 11 (1995).
\bibitem{Kurdyubov-FTT2017} A. S. Kurdyubov, A. V. Trifonov, I. Ya. Gerlovin, I. V. Ignatiev, and A. V. Kavokin, Phys. Solid St. {\bf 59}, 1298 (2017) [Fiz. Tv. Tela {\bf 59}, 1274 (2017)].
\bibitem{Dumke-PR1963} W. P. Dumke, Phys. Rev. {\bf 132}, 1998 (1963).
\bibitem{Wang-APL2017} G. Wang, B. Zhang, H. Ji, X. Liu, T. He, L. Lv, Y. Hou, and J. Shen, Appl. Phys. Lett. {\bf 110}, 023301 (2017).
\bibitem{Matsui-OptLett2013} T. Matsui, R. Takagi, K. Takano, and M. Hangyo, Opt. Lett. {\bf 38}, 4632 (2013).
\bibitem{Sze-1968} S. M. Sze, J. C. Irvin, Solid State Electron. {\bf 11}, 599 (1968).
\bibitem{Schultz-model2009} P. A. Schultz and O. A. von Lilienfeld, Modelling Simul. Mater. Sci. Eng. {\bf 17}, 084007 (2009).
\bibitem{Heilman-SST1990} R. Heilman and G. Oelgart, Semicond. Sci. Technol. {\bf 5}, 1040-1045 (1990).
\bibitem{Strauch-JPhys1990} D. Strauch and B. Dorner, J. Phys.: Condens. Matter. {\bf 2}, 1457 (1990).

\end{thebibliography}
\end{document}